# On the Approximate Analysis of Energy Detection over $n$*Rayleigh Fading Channels through Cooperative Spectrum Sensing

Yahia Alghorani, *Student Member, IEEE*, Georges Kaddoum, *Member, IEEE*, Sami Muhaidat, *Senior Member, IEEE*, and Samuel Pierre, *Senior Member, IEEE*

*Abstract*—In this letter, we consider the problem of energy detection of unknown signals in an intervehicular communication (IVC) system over $n$*Rayleigh fading channels (also known as cascaded Rayleigh). Novel tight approximations for the probability of detection are derived for the no-diversity and the maximum ratio combining (MRC)) diversity schemes. Moreover, we investigate the system performance when cooperative spectrum sensing (CSS) is considered with and without imperfect reporting channels. The analytical results show that the detection reliability is decreased as the fading severity parameter $n$ increases and that reliability is substantially improved when CSS employs MRC schemes.

*Index Terms*—Cognitive radio, energy detection, $n$*Rayleigh fading channels, diversity schemes, cooperative spectrum sensing.

## I. INTRODUCTION

R ecent studies have reported the importance of using cognitive radio in vehicular networks as a candidate solution for spectrum scarcity, and various techniques have been studied for cooperative spectrum sensing (CSS) as a one way to enhance local spectrum sensing [1, 2]. However, one of the most challenges of implementing spectrum sensing in vehicular networks is the hidden terminal problem which occurs when the CR (secondary) is hidden in severe multipath fading while the primary user (PU) is operating in the vicinity [3]. Such kind of problems can cause interference to the licensed users due to the very low SNR environments, which in turn leads to degrade the system performance. Hence, cooperative spectrum sensing has interestingly become a powerful solution to increase the probability of PU detection [4]. In this case, CR users have ability to collect radio environmental information such as power, position, frequency and bandwidth, then report it to fusion center (FC) where the local observations from CR users, are fused to detect the presence of PU. There are several factors play an important role in PU detection process, for instance, the position of CR user as a relay, whether it is located nearby or far away from the PU. By assuming the CR user is nearby the PU, the signal can be detected reliably with short sensing duration, which in turn makes the whole sensing time short at the FC. In addition, the number of cooperative partners is also another factor. By increasing the number of CR users, the probability of detecting the PU can be improved significantly but it makes the whole sensing time long [5].

It has been demonstrated recently that the so-called $n$*Rayleigh fading channel model, which involves the product of two or more independent Rayleigh distributed random variables, provides an accurate statistical description of the intervehicular communications (IVC) channels [6]. In [7, 8], experimental results in different vehicular communication environments have shown that in vehicular networks, several small-scale fading processes are multiplied together, leading to a worse-than Rayleigh fading. Unfortunately, most of the studies in the literature discussed cooperative spectrum sensing with energy detection over classical Rayleigh fading channels, which is limited to fixed-to-mobile cellular radio channels (see, e.g, [9, 10] and reference therein). There are very few studies that have discussed the problem of energy detection over $n$*Rayleigh distribution, which is presented as an accurate statistical propagation model for mobile-to-mobile scenario. For instance, in [11], the detection probability over double Rayleigh fading channels (i.e., $n = 2$) has been investigated with and without receive diversity reception such as selection combining (SC) and switch and stay combining (SSC). The results have shown that the energy detection over double Rayleigh fading is worse than that of Rayleigh channels, and is improved by diversity systems.

In this letter, we analyze the energy detection over $n$*Rayleigh fading channels (i.e, $n > 2$) when CSS scheme is employed in IVC system with and without maximum ratio combining (MRC) diversity reception. To the best of our knowledge, has not been studied yet. Specifically, each local detector (CR user) makes a local binary decision, depending on whether the CR user decides $H_0$ (PU is absent) or $H_1$ (PU is present). All local decisions are then reported to a common receiver-enabled CR (i.e., a moving vehicle with a data fusion center), where they are combined to make a final decision about the presence or the absence of the PU. Through this approach, we analyze the spectrum sensing performance under various detection constraints, in cases of perfect reporting channels (where the reporting channel is free of error) and imperfect reporting channels (where the reporting channels between the CR users and the common receiver suffer from severe fading). Accordingly, approximate closed-form expressions of the average detection probability for both the

Y.Alghorani and S.Pierre are with the Mobile Computing and Networking Research Laboratory (LARIM), Department of Computer Engineering, Ecole Polytechnique de Montreal, P.O. Box 6079, Station Centre-ville, Montreal, Quebec, H3C 3A7, Canada (e-mail: yahia.alghorani@polymtl.ca; samuel.pierre}@polymtl.ca).

G.Kaddoum is with the Electrical Engineering Department, École de Technologie Supérieure, Montréal, Quebec, Canada, H3C1K3, (email: Georges.Kaddoum@etsmtl.ca).

S. Muhaidat is with the Electrical and Computer Engineering Department, University of Western Ontario, London, Ontario, N6A 5B9 , Canada, (e-mail: sami.muhaidat@uwo.ca).

no-diversity and diversity reception cases $\overline{P_{dt}^n}$ are presented, and the receiver operating characteristic (ROC) curves (probability of missed detection $Q_m = 1 - Q_d$ versus probability of false alarm $Q_f$ ) are obtained for different values of the fading severity parameter $n$.

The rest of this letter is organized as follows: in Section II, the performance of local spectrum sensing over $n$*Rayleigh fading channels is introduced. Section III studies the impact of diversity reception on $n$*Rayleigh fading channels. In Section IV, cooperative spectrum sensing is investigated. In Section V, the numerical evaluation of cooperative spectrum sensing with and without imperfect reporting channels is presented. Finally, conclusions are given in Section VI.

## II. LOCAL SPECTRUM SENSING WITH NO-DIVERSITY

We consider a vehicular network composed of $M$ CR users equipped with a single pair of transmit and receive antennas. Each CR user performs its spectrum sensing independently to decide between the following two hypotheses

$$x_i(t) = \begin{cases} w_i(t), & : H_0 \\ h_i(t)s(t) + w_i(t), & : H_1 \end{cases} \quad (1)$$

where $x_i(t)$ is the signal received by the $i$th CR user in time slot $t$, $s(t)$ is the PU transmitted signal, $w_i(t)$ is the additive white Gaussian noise (AWGN) with zero mean and variance ($N_o W$)-denoted by one-sided noise and signal bandwidth, and $h_i(t)$ is a product of $n$ independent circularly complex Gaussian random variables of the sensing channel between the PU and the $i$th CR user, which is defined as $h_i \triangleq \prod_{j=1}^{n} h_{i,j}$ with zero mean and channel variance of ($\eta_i$), for $i = 1, ..., M$. Hence, $|h_i|$ follows $n$*Rayleigh distribution. We assume the received signal undergoes slow fading (i.e., the symbol period of the detected signal is smaller than the coherence time of the channel), which can be justified for rush-hour traffic. At each CR user, the received signal is filtered, squared and integrated over a time interval $T$, and the energy of output signal measured by $y_i \triangleq \frac{2}{N_o} \int_0^T |x_i(t)|^2 dt$ [12]. Therefore, $y_i$ acts as a test statistic and follows a central chi-square distribution with $2TW$ degrees of freedom under $H_0$ and a non-central chi-square distribution with $2TW$ degrees of freedom under $H_1$.

In order to find an approximate expression for the average probability of detection over $n$*Rayleigh fading channels $\overline{P_{dt}^n}$, we need to recall the probabilities of false alarm and detection for a non-fading AWGN channel, which are given respectively by [12]

$$P_{fi} = P\{y_i > \lambda_i | H_0\} = \frac{\Gamma(u, \lambda_i/2)}{\Gamma(u)}, \quad (2)$$

$$P_{di} = P\{y_i > \lambda_i | H_1\} = Q_u\left(\sqrt{2\gamma_i}, \sqrt{\lambda_i}\right). \quad (3)$$

where $u = TW$, $\lambda_i$ and $\gamma_i = |h_i|^2 E_s/N_o W$ are the energy detection threshold and the instantaneous signal-to-noise ratio (SNR) at the $i$th CR user respectively. $E_s$ is the transmission energy per symbol, $\Gamma(.)$ is the gamma function defined by $\Gamma(\alpha) = \Gamma(\alpha, 0)$ [13], $\Gamma(.,.)$ represents the upper incomplete gamma function, defined by $\Gamma(\alpha, x) = \int_x^\infty e^{-t} t^{\alpha-1} dt$ [13], and $Q_u(.,.)$ is the generalized Marcum-Q function defined as $Q_u(a,b) = \int_b^\infty \frac{x^u}{a^{u-1}} e^{-\frac{x^2+a^2}{2}} I_{u-1}(ax)dx$ [14], where $I_{u-1}(.)$ is the modified Bessel function of the first kind and order $u - 1$.

The generalized Marcum Q-function can be expressed in the infinite series representation as [15, eq. (29)]. To converge more quickly, by having a finite series, an asymptotic solution for the generalized Marcum Q-function has been introduced in [16, eq.(6)] for arbitrary values of its order $u$, which is given in a finite series representation and leads us to rewrite (3) in an approximate expression as

$$P_{di} \approx \sum_{l=0}^{k} \frac{\Gamma(k+l)k^{1-2l}\Gamma\left(u+l, \frac{\lambda_i}{2}\right)\gamma_i^l}{\Gamma(l+1)\Gamma(k-l+1)\Gamma(u+l)e^{\gamma_i}}. \quad (4)$$

Since $k = \infty$ is exact, it's worthwhile to mention that $k$-value selection in (4) is related to the involved parameters (i.e., $u$, $\lambda_i$, $\gamma_i$). Thus, the accuracy of (4) depends on the value of $k$ and improves when $k$ increases [16].

Under $n$*Rayleigh distribution [17] with the help of [18, eq.(2.3)], the exact probability density function (PDF) of instantaneous SNR $\gamma_i$ at the $i$th CR user is given by

$$f_{\gamma_i}(\gamma_i) = \sqrt{2^{1-n_i}} \ (\bar{\gamma}_i\gamma_i)^{-\frac{1}{2}} G_{0,n_i}^{n_i,0}\left(2^{1-n_i}\frac{\gamma_i}{\bar{\gamma}_i} \bigg| \frac{1}{2}, ..., \frac{1}{2}\right). \quad (5)$$

where $G_{p,q}^{m,n}(.)$ is the Meijer-G function defined in [13, eq.(9.301)], and $\bar{\gamma}_i = \mathbf{E}(|h_i|^2)E_s/N_o W$ is the average SNR of $\gamma_i$, where $\mathbf{E}(.)$ is the statistical expectation operator. By averaging (4) over (5) and using the fact defined in [19, eq.(07.34.21.0088.01)], the average probability of detection for the $i$th CR user over $n$*Rayleigh fading channels can be obtained as

$$\overline{P_{dt}^n} \approx \sum_{l=0}^{k} \frac{\Gamma(k+l)k^{1-2l}\Gamma\left(u+l, \frac{\lambda_i}{2}\right)}{\Gamma(l+1)\Gamma(k-l+1)\Gamma(u+l)}$$

$$\times \sqrt{2^{1-n_i}} \ \bar{\gamma}_i^{-\frac{1}{2}} G_{1,n_i}^{n_i,1}\left(\frac{2^{1-n_i}}{\bar{\gamma}_i} \bigg| \begin{matrix} \frac{1}{2}-l \\ \frac{1}{2}, ..., \frac{1}{2} \end{matrix}\right). \quad (6)$$

However, a new accurate approximation for the PDF of $n$*Rayleigh distribution has been proposed in [20], which is derived based-on a transformed Nakagami-$m$ distribution as

$$f_{\gamma_i}(\gamma_i) \approx \frac{\beta_i^{m_i}\gamma_i^{\alpha_i-1}}{n_i\Gamma(m_i)} e^{-\beta_i\gamma_i^{\frac{1}{n_i}}}, \quad \gamma_i \geq 0 \quad (7)$$

where $\alpha_i = m_i/n_i$ and $\beta_i = m_i/\Omega_i\bar{\gamma}_i^{1/n_i}$. The fading severity parameters ($m_i, \Omega_i$) are given by [20]

$$m_i = 0.6102n_i + 0.4263, \quad \Omega_i = 0.8808n_i^{-0.9661} + 1.12$$

It's important to mention that the novel approximation for the PDF in (7) has been examined in [20], by comparing it to the exact PDF derived in (5). The results have shown that the new approximation has high accuracy in most cases considered. Furthermore, the approximate PDF is easy to calculate and to manipulate compared to the exact PDF as noticed later for deriving the PDF of the SNR at the output of MRC scheme.

By averaging (4) over (7) and using the fact that $\exp(-x) = G_{0,1}^{1,0}\left(x \mid \begin{smallmatrix} - \\ 0 \end{smallmatrix}\right), \forall x$ [19], the approximate closed-form expression for the average probability of detection for the $i$th CR user can be found as

$$\overline{P_{dt}^n} \approx \sum_{l=0}^{k} \frac{\Gamma(k+l)k^{1-2l}\Gamma\left(u+l,\frac{\lambda_i}{2}\right)}{\sqrt{n_i}\,\Gamma(m_i)\Gamma(l+1)\Gamma(k-l+1)\Gamma(u+l)}$$

$$\times \frac{\beta_i^{m_i}}{(2\pi)^{\frac{n_i-1}{2}}} G_{1,n_i}^{n_i,1}\left(\left(\frac{\beta_i}{n_i}\right)^{n_i} \middle| \begin{matrix} 1-(\alpha_i+l) \\ 0,\ldots,\frac{n_i-1}{n_i} \end{matrix}\right). \quad (8)$$

The Meijer-G function in (8) can be evaluated much faster than that of (6) using the common mathematical software packages such as MATHEMATICA or MAPLE. This is due to the fact that (8) is built based on a simple PDF compared to (6) which requires high computational complexity of the PDF in (5).

### III. LOCAL SPECTRUM SENSING WITH DIVERSITY RECEPTION

In this section, we derive an approximate expression for the average probability of detection when the MRC diversity reception is employed at each CR user. In such a scenario, the diversity scheme is employed to combat the severe fading due to $n^*$Rayleigh fading channels. We consider $L$ diversity branches which are independent and identically distributed (i.i.d), and are subject to $n^*$Rayleigh fading. The instantaneous SNR of the combiner output for $i$th CR user is given by $\gamma_{ti} = \sum_{r=1}^{L} \gamma_{i,r}$ [18].

For a non-fading AWGN channel, the probabilities of false alarm and detection at the output of MRC are the same as (2) and (3), respectively. The PDF of $\gamma_{ti}$ for the $i$th CR user over i.i.d $n^*$Rayleigh fading random variables can be derived by normalizing (7) to the PDF of the SNR at the MRC output with Nakagami fading [21, eq.(7)], to be finally expressed as

$$f_{\gamma_{ti}}(\gamma_{ti}) \approx \frac{(L\beta_i)^{Lm_i}\gamma_{ti}^{L\alpha_i-1}}{n_i\Gamma(Lm_i)L^{L\alpha_i}} \exp\left(-\frac{L\beta_i}{L^{1/n_i}}\gamma_{ti}^{\frac{1}{n_i}}\right). \quad (9)$$

The PDF of $\gamma_{ti}$ is quite similar to that in (7) by replacing $m_i$ by $Lm_i$ and $\overline{\gamma}_i$ by $L\overline{\gamma}_i$. However, eq.(9) is novel and has not been derived yet for the MRC scheme with $n^*$Rayleigh fading. Similarly, the average detection probability for the MRC diversity scheme at the $i$th CR user, can be evaluated by averaging $P_{di}$ over (9).

$$\overline{P_{dt}^n} \approx \sum_{l=0}^{k} \frac{\Gamma(k+l)k^{1-2l}\Gamma\left(u+l,\frac{\lambda_i}{2}\right)}{\sqrt{n_i}\,\Gamma(Lm_i)\Gamma(l+1)\Gamma(k-l+1)\Gamma(u+l)}$$

$$\times \frac{(L\beta_i)^{Lm_i}}{(2\pi)^{\frac{n_i-1}{2}}} G_{1,n_i}^{n_i,1}\left(\left(\frac{L\beta_i}{n_i}\right)^{n_i} \middle| \begin{matrix} 1-(L\alpha_i+l) \\ 0,\ldots,\frac{n_i-1}{n_i} \end{matrix}\right). \quad (10)$$

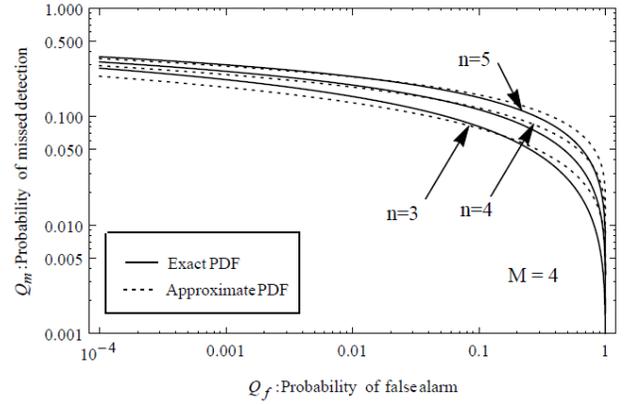

**Fig. 1.** Comparison of complementary ROC curves for cooperative spectrum sensing-based the exact PDF and approximate PDF in (7) with different $n$ values and perfect reporting channels ($\overline{\gamma} = 10$dB, $u = 5$, $(n = 3; \ m = 2.25, \ \Omega = 1.42), (n = 4; \ m = 2.86, \ \Omega = 1.35), (n = 5; \ m = 3.47, \ \Omega = 1.30)$).

### IV. COOPERATIVE SPECTRUM SENSING

In cooperative spectrum sensing, each CR user sends its binary decision $D_i \in \{0,1\}$, to fusion center (FC). At the FC, all the local decisions are combined according to the OR fusion rule which gives better performance than other rules [22]. Consequently, the global decision is made to infer the presence or the absence of PU after $M$ symbol periods. In this case, we assume the instantaneous SNRs at the CR users, are independent, but not necessarily identically (i.n.i.d) distributed random variables. In addition, all CR users use different decision thresholds $\lambda_i$, for $i = 1,\ldots,M$. Hence, the global probabilities of false and miss through an error-free channel, are given respectively by [23]

$$Q_f = 1 - \prod_{i=1}^{M}(1 - P_{fi}), \quad Q_m = \prod_{i=1}^{M}\overline{P_{mi}^n}. \quad (11)$$

where $\overline{P_{mi}^n} = 1 - \overline{P_{di}^n}$ is the average probability of missed detection of the local spectrum sensing of the $i$th CR user.

On the other hand, in case of imperfect reporting channels between $i$th CR user and the FC, the global probabilities of false alarm and miss are given respectively by [24]

$$Q_f = 1 - \prod_{i=1}^{M}[(1 - P_{fi})(1 - P_{ei}) + P_{fi}P_{ei}], \quad (12)$$

$$Q_m = \prod_{i=1}^{M}\left[\overline{P_{mi}^n}(1 - P_{ei}) + \left(1 - \overline{P_{mi}^n}\right)P_{ei}\right]. \quad (13)$$

where $P_{ei}$ is the probability of reporting error between the $i$th CR user and the common receiver.

### V. NUMERICAL RESULTS

In this section, we analyze the performance of cooperative spectrum sensing with and without diversity reception over $n^*$Rayleigh fading channels. The performance analysis has been implemented under various channel constraints.

Fig.1 shows the cooperative spectrum sensing performance without diversity reception over $n^*$Rayleigh fading channels.

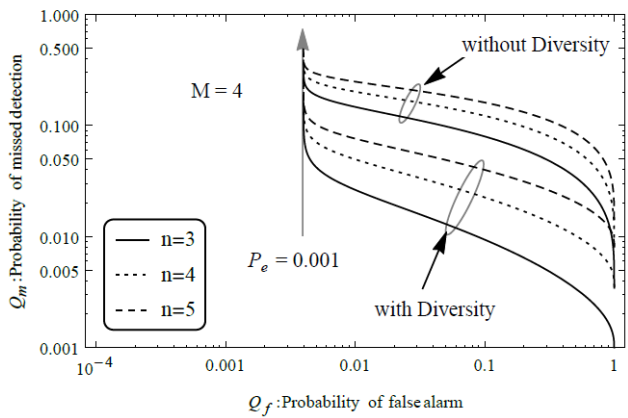

**Fig.2.** Complementary ROC curves for cooperative spectrum sensing with MRC diversity reception over $n$*Rayleigh fading and imperfect reporting channels ($L = 3$, $\bar{\gamma} = 10$dB, $u = 5$).

The figure compares the performance results-based the exact PDF (5) with that of the approximate PDF (7) under perfect reporting channels. In this case, we set $k = 500$ and the accuracy of analysis improves as the value of $k$ increases. As observed from Fig.1, the detection performance-based the approximate PDF is closely fits the analytical results-based the exact PDF. The larger the value of $n$ is, the higher accuracy will be [20]. Furthermore, we can see that, increasing $n$ would result in missing the presence of the primary user, which leads to increase the interference to licensed users. Therefore, CSS is more attractive in practical scenarios as $n$ increases.

Fig.2 shows the effect of MRC diversity scheme on the cooperative spectrum sensing performance over $n$*Rayleigh fading channels. In this scenario, we analyze the system performance under imperfect reporting channels. As clearly observed from this figure, MRC scheme has a positive impact on the probability of missed detection $Q_m$ compared to the no-diversity system. In particular, there is an obvious diversity gain, about one order of magnitude for all values of $n$, and this gain starts to decrease as $n$ increases. Further, when the false alarm probability reaches to a threshold (i.e, $Q_f \geq MP_e$) [24], the probability of missed detection $Q_m$ tends to one, which leads to deteriorate the detection performance. In general, MRC diversity scheme can bring additional gain to CSS when the probability of false alarm $Q_f$ is considered, in such a way that CSS keeps the $Q_f$ minimal rather than increasing the number of cooperative users, which results in increasing the probability of false alarm and low spectrum utilization as observed in [24].

## VI. CONCLUSION

In this letter, we have studied the performance of an energy detector over $n$*Rayleigh fading channels. Approximate expressions of the average detection probability are derived for both the no-diversity and receive diversity schemes. The performance of cooperative spectrum sensing has been further investigated with and without imperfect reporting channels. The results indicate that the probability of detection deteriorates when the fading severity parameter $n$ increases.

Finally, we have demonstrated that the underlying CSS scheme with diversity reception improves the sensing diversity order but this improvement is limited under imperfect reporting channels.